\documentclass[aps,pre,showpacs,showkeys,twocolumn,floatfix,superscriptaddress]{revtex4-1}

\usepackage{amssymb}
\usepackage{amsmath}
\usepackage{graphicx}
\usepackage{amsbsy}
\usepackage{color}
\usepackage{bm}

\newcommand{\bq}{\begin{eqnarray}}
\newcommand{\eq}{\end{eqnarray}}
\newcommand{\bqn}{\begin{eqnarray*}}
\newcommand{\eqn}{\end{eqnarray*}}

\begin{document}
\title{From toroidal to rod-like condensates of semiflexible polymers}

\author{Trinh Xuan Hoang}
\email{hoang@iop.vast.ac.vn}
\affiliation{Center for Computational Physics,
Institute of Physics, Vietnam Academy of Science and Technology,
10 Dao Tan St., Hanoi, Vietnam}
\author{Achille Giacometti}
\email{achille.giacometti@unive.it}
\affiliation{Dipartimento di Scienze Molecolari e Nanosistemi, Universita' Ca' Foscari Venezia, I-30123 Venezia, Italy}
\author{Rudolf Podgornik}
\email{rudolf.podgornik@ijs.sl}
\affiliation{Department of Theoretical Physics, J. Stefan Institute and Department of Physics, Faculty of Mathematics and Physics, University of Ljubljana - SI-1000 Ljubljana, Slovenia, EU}
\affiliation{Department of Physics, Faculty of Mathematics and Physics, University of Ljubljana, SI-1000 Ljubljana, Slovenia, EU}
\affiliation{Department of Physics, University of Massachusetts, Amherst, MA 01003, USA}
\author{Nhung T.T. Nguyen}
\email{ntnhung@iop.vast.ac.vn}
\affiliation{Center for Computational Physics,
Institute of Physics, Vietnam Academy of Science and Technology,
10 Dao Tan St., Hanoi, Vietnam}
\author{Jayanth R. Banavar}
\email{banavar@umd.edu}
\affiliation{Department of Physics, University of Maryland, Collega Park, Maryland 20742, USA}
\author{Amos Maritan}
\affiliation{Dipartimento di Fisica, Universit\`{a} di Padova, via
Marzolo 8 I-35131 Padova}
\affiliation{CNISM, Unit\`a di Padova, Via Marzolo 8, I-35131 Padova, Italy}
\affiliation{Sezione INFN, Universit\`a di Padova, I-35131 Padova, Italy}
\email{maritan@pd.infn.it}

\date{\today}

\begin{abstract}
The competition between toroidal and rod-like conformations as possible ground states for DNA condensation is studied as a function of the stiffness, the length of the DNA and the form of the long-range interactions between neighboring molecules, using analytical theory supported by Monte Carlo simulations. Both conformations considered are characterized by a local nematic order with hexagonal packing symmetry of neighboring DNA molecules, but differ in global configuration of the chain and the distribution of its curvature as it wraps around to form a condensate.  
{The long-range interactions driving the DNA condensation are assumed to be of the form pertaining to the attractive depletion potential as well as the attractive counterion induced soft potential.}
In the stiffness-length plane we find a transition between rod-like to toroid condensate for increasing stiffness at a fixed chain length $L$. Strikingly, the transition line is found to have a $L^{1/3}$ dependence irrespective of the details of the long-range interactions between neighboring molecules. When realistic DNA parameters are used, our description reproduces rather well some of the experimental features observed in DNA condensates.
\end{abstract}

\maketitle

\section{Introduction}
Double-stranded DNA (dsDNA) is a linear semiflexible polymer chain with persistence lengths of about 150 base pairs (bp) ($50$ nm) and cross-section diameter of about $2$ nm. In aqueous solutions DNA molecules are highly negatively charged due to dissociated phosphate groups along the chain backbone that confer to B-form ds-DNA a bare base-pair charge of $2 e_0$ per 0.34 nm length of DNA, engendering strong repulsive interactions along and between DNA molecules \cite{RP1}. Nevertheless, under specific solution conditions DNA molecules can be induced to condense into highly compact structures that phase separate from the solution \cite{DNAcond}. In these condensates DNA is in a liquid crystalline state \cite{Maniatis,Evdokhimov} with lattice spacings close to measured spacings in bulk DNA liquid crystals at the same solution conditions \cite{Rau84,Podgornik94,Strey99}. The phenomenon of semiflexible polymer condensation is not specific to DNA only but can be observed in other semiflexible polyelectrolytes as well, e.g. F-actin filaments \cite{Tang}. The morphology of these condensates varies depending on the method of preparation \cite{Hud05} as well as on the dynamics of the nucleation and growth of the condensate \cite{Vilfan06,Muthukumar2005}. In what follows we will make the simplifying assumption that the condensate morphology is an equilibrium property, eventually reached after different nucleation and growth relaxation processes are over, and study the consequences.

Previous attempts at a theoretical analysis of DNA condensation in the presence of various condensing agents shed light on the resulting ordered nematic DNA structure often in the form of toroidal and/or rodlike globules \cite{Gros_khok}. Different aspects of this collapse transition have been scrutinized in order to deduce the detailed geometry of the aggregate and the corresponding phase diagram \cite{history} as well as their dependence on the assumed form of the elastic energy and the DNA-DNA or DNA-condensing agent interaction \cite{Interact}. In what follows we will revisit the problem of the stability and phase diagram of the various condensed structures of DNA in light of the recent understanding of the interactions driving the condensation transition in the case of polyvalent counterions and osmoticants, as well as the emerging details of the non-linear nature of the DNA elastic energy. The theoretical approach advocated  here, combined with Monte Carlo simulations, provides a simple and unified foundation on which the effects of various components of the DNA condensation phenomenon can be compared and assessed.

When condensed in a very dilute solution,  the most commonly observed DNA condensate morphologies are torus-like and rod-like (see Figs.\ref{fig:fig1} and \ref{fig:fig2}). A typical compact structure has the size of approximatively $100$ nm, with an inner hole about $30$ nm wide in the case of the toroidal aggregate. It is found to be relatively robust with respect to the length of the DNA involved in the condensation. The packing of DNA strands inside the condensed structure is highly ordered with a predominantly hexagonal packing in the plane perpendicular to the toroidal main axis \cite{Rawat09,Stukan06, Ishimoto08,Stukan03, Schnurr02, Starostin13}. The condensation can be induced by a variety of {\sl condensing agents}.
Among these flexible polymers, such as PEG (poly-ethylene-glycol), at large enough concentrations, the presence of salt (PSI-condensation = (P)olymer and (S)alt (I)nduced condensation) induces condensation of both DNA as well as F-actin filaments \cite{Interact,Renko}.

 The mechanism here appears to be osmotic depletion interactions \cite{Depletion} due to the exclusion of the polymer from the DNA subphase. More commonly exploited condensation agents in various biological settings are the multivalent counterions. In fact many, but not all, multivalent cations induce ds-DNA condensation. Those that do condense ds-DNA at finite concentrations are Mn$^{2+}$, Cd$^{2+}$, Co(NH$_3$)$^{3+}$, polyamines such as spermidine$^{3+}$, spermine$^{4+}$, polylysine$^{+}$ and all the higher valency (poly)counterions. That electrostatics plays an important role in DNA condensation is clear but it is just as clear that it can not be the only factor driving it \cite{Curop,Cherstvy2011,Teif}. Furthermore, a radical reformulation of the theory of electrostatic interactions is needed \cite{Rudi2010,Rudibook}, based on the concept of the "strong-coupling" electrostatics between the multivalent salt counterions and the charges on the DNA backbone, in order to understand the counterintuitive change in sign of interactions between nominally equally charged bodies \cite{Rudibook,Perspective}.

While electrostatics should play an important role in the DNA condensation mechanism, it cannot be the sole and sometimes not even the dominant factor affecting it. For instance $\mathrm{Co(NH_3)}^{3+}$ is more efficient in condensing DNA than spermidine$^{3+}$, both being trivalent counterions, and the best condensing agents appear to be those that bind into one of the DNA grooves \cite{DeRouchey10}. These well documented ion specific effects \cite{Rau92}  furthermore suggest that interaction of condensing ions with water molecules, i.e. hydration interactions, provide the necessary specificity that is absent in condensation interactions based exclusively on Coulomb interaction [for a recent review of hydration effects see Ref \onlinecite{Rau11}].

Apart from the polymer depletion and "strong-coupling" electrostatic interactions, a fundamental ingredient of any theory of semiflexible polymer condensation is their significant stiffness that frustrates the formation of a spherical globule \cite{Yamakawa,Svensek}. Without any stiffness effects, one would expect that the spherical globule would be the ground state of a flexible polymer by minimizing its surface energy \cite{Grosbergbook}. Indeed, the local structure of this condensed phase consists of straight chains with parallel nematic alignment to minimize the bending energy. 
However, even within this simple picture, it is not clear why this particular structure is necessarily favored with respect to other structures – such as, for instance, a rod-like structure – having similar characteristics. Nor is the role of the concrete form of the elastic energy as well as the interactions between DNA molecules that induce the condensation well understood. 


In this paper we will address these issues using simple analytical arguments supported by numerical simulations, and discuss under what conditions the toroidal condensate is favored with respect to a rod-like counterpart. We will include the bending energy, the surface energy and the detailed interaction energy between polymer molecules in our {\sl {\sl Ansatz}} for the free energy whose minimization will provide us with the  equilibrium configuration of the condensate.

The outline of the paper is as follows: we first present the model with the corresponding packing geometry, and the non-equilibrium curvature, surface and interaction free energies written for the dominant configurations of a toroid and a spherocylinder. We then proceed to the analytical minimization of the total free energy for the toroidal and rod-like aggregate that we compare with Monte-Carlo simulations. We finally explore the effect of various interaction models and generalize the elastic energy {\sl {\sl Ansatz}} to the case of an intrinsic threshold. We conclude with a commentary on previous works and with an assessment on the validity of our approach.

\section{Model}

We consider a semiflexible polymer of length $L$ formed by $N$ spherical beads of diameter $b$. The bond length between consecutive beads is also taken to be equal to $b$ for simplicity.  A conformation of the polymer is given by the positions of the beads $\{{\bf r}_i,i=1,2,\ldots N\}$. We furthermore assume that all the energies involved in condensate morphology are sufficiently large so that entropic terms can be neglected. Hence we expect a compact phase for which only curvature, surface and interaction energy terms are present.

\subsection{Packing geometry}

In our model the polymer chain fills the condensate interior and is locally hexagonally packed. Consider a tessellation of a plane perpendicular to the long axes of the polymer, having a Schl\"afli symbol \cite{Coxeter69} ${p,q}$. For hexagonal packing the Schl\"afli symbol is ${6,3}$. The hexagonal packing fraction \cite{Coxeter69} 
\begin{eqnarray}
\label{sec1:eq1}
\eta &=& \frac{\pi}{p} \cot\left(\frac{\pi}{p}\right)
\end{eqnarray}
for the case of hexagonal close packing ($p=6$) is then given by
\begin{eqnarray}
\eta_{\text{hex}}&=& \frac{\pi}{6}\cot\left(\frac{\pi}{6}\right) = \frac{\pi}{2 \sqrt{3}} =0.9069...
\label{sec1:eq2}
\end{eqnarray}
If the packing of polymers inside the aggregate is still hexagonal but not at the highest close packing fraction $\eta_{\text{hex}}$, where the separation between the polymer chains is equal to $d$, then $d \ge b$ and the packing fraction is given by:
\begin{equation}
\label{sec1:eq3}
\eta = \eta_{\text{hex}} \left(\frac{b}{d}\right)^2 .
\end{equation}
The hexagonal packing symmetry is thus still preserved but at a lower packing fraction. This will be important when we introduce soft interactions between polymer segments.

\subsection{Curvature and surface energy}
\label{subsec:curvature}

Within the worm-like chain model \cite{Rubinstein03} the elastic bending free energy is given by  
\begin{equation}
\label{sec2:eq1}
U = \kappa \sum_{i=2}^{N-1} (1-\cos \theta_i),
\end{equation}
where $\kappa$ is the (reduced) stiffness, $\theta_i$ is the angle between ${\bf r}_{i-1,i}$
and ${\bf r}_{i,i+1}$ with ${\bf r}_{ij}={\bf r}_j - {\bf r}_i$. The energies are assumed to be in units of thermal energy $k_B T$, so that both $U$ and $\kappa$ are dimensionless variables. Here $k_B$ is the Boltzmann constant, $T$ the absolute temperature, and  we denote as $l_p = \kappa b$ the persistence length of the chain. If $R_i$ is the radius of curvature at the bead $i$, i.e. the radius of the circle through
${\bf r}_{i-1}$, ${\bf r}_{i}$, and ${\bf r}_{i+1}$, it is easy to show that
\begin{equation}
\label{sec2:eq2}
1 - \cos \theta_i = \frac{b^2}{2 R^2_i} \ .
\end{equation}

\begin{figure}[htbp]
\begin{center}
   \includegraphics[width=2.0in,angle=270]{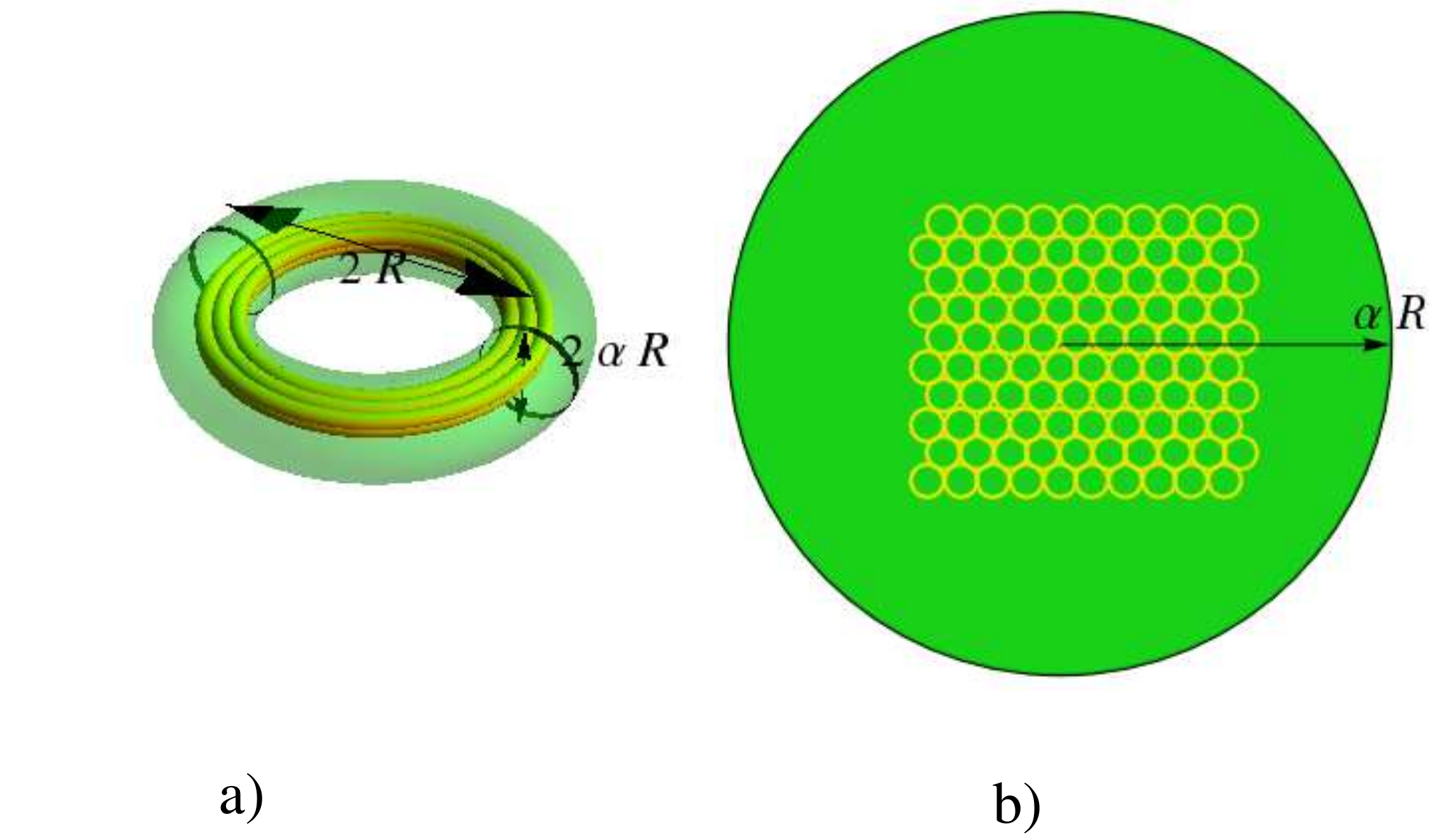} 
\end{center}  
\caption{(Left) Sketch of the DNA condensate as a toroidal phase. The mean
radius of the toroid is $R$. The radius of cross section is $\Delta = \alpha R$
with $0<\alpha<1$. (Right) The hexagonal packing within the cross section of
the torus.}
\label{fig:fig1}
\end{figure}

Assume that the chain forms a toroid of the mean radius $R$ and radius of the cross-section
$\Delta=\alpha R$ with $0 < \alpha < 1$ (see Fig. \ref{fig:fig1}). The limiting case $\alpha=0$ ($R \to \infty$) would correspond to a swollen conformation whereas $\alpha=1$ would correspond to a ``globular'' conformation with no inner hole inside the torus (not a sphere).

In the simplest case, one can furthermore  assume that the chain has a constant radius of curvature equal to $R$. 
The bending energy in Eq. (\ref{sec2:eq1}) in the large $N$ limit
then simplifies to
\begin{equation}
U_{toroid} = \frac{\kappa}{2} \frac{Lb}{R^2} \ ,
\label{eq:etor1}
\end{equation}
where $L=Nb$.
Note that our constant radius of curvature approximation follows
the model proposed in Ref. \onlinecite{Hud95}
which claimed to predict a correct toroid size distribution. This
approximation may not be fully consistent with our previous assumption
on the hexagonal packing. Thus, we assume that the latter is not rigorously
valid but only essentially correct. An alternative model of DNA organization
inside the condensates based on perfect hexagonal packing \cite{Starostin06}  yields similar expression for bending
energy but only in the limit of thin condensate.
We assume (see Fig.\ref{fig:fig1} (b)) that the polymer tightly wraps around
the torus with $n_{loops}$ loops that are related to the two-dimensional
(cross-section) packing fraction $\eta$ given in Eq.
(\ref{sec1:eq2}) by the ratio of the occupied to the total surface
\begin{equation}
\label{sec2:eq3}
\eta = \frac{n_{loops} \pi (b/2)^2}{\pi (\alpha R)^2} .
\end{equation}
In order to compare low-energy configurations with different geometries, we need to translate the typical length scale of the problem ($R$ in this case) to the contour length $L$ of the polymer that is common to all configurations.   In the present case, we have
\begin{equation}
\label{sec2:eq5}
L= 2 \pi R n_{loops} = 8 \pi R  \eta \alpha^2 \left(\frac{R}{b} \right)^2 
\end{equation} 
Thus we end up with
\begin{equation}
\frac{R}{b} = \left(8 \pi \eta \alpha^2\right)^{-1/3} \left(\frac{L}{b} \right)^{1/3}
\label{eq:Rb}
\end{equation}
Inserting Eq. (\ref{eq:Rb}) into Eq. (\ref{eq:etor1}),  one obtains
\begin{equation}
U_{toroid} =  2\kappa \pi^{2/3} \eta^{2/3} \alpha^{4/3}
\left(\frac{L}{b}\right)^{1/3} .
\label{sec2:eq6}
\end{equation}

On the other hand, if we denote as $\sigma$ the surface tension, the surface energy of the toroid is, using Eq.(\ref{eq:Rb}) again,       
\begin{eqnarray}
\sigma S_{toroid} &=& \sigma (2\pi\alpha R)(2\pi R) =
\nonumber \\
&=& \left(\sigma b^2\right) \pi^{4/3} \eta^{-2/3}\alpha^{-1/3}  \left(\frac{L}{b}\right)^{2/3}.
\label{eq:storoid}
\end{eqnarray}
Note that
\begin{equation}
U_{toroid} \sim \left(\frac{L}{b} \right)^{1/3} \qquad \sigma S_{toroid} \sim  \left(\frac{L}{b} \right)^{2/3}
\end{equation}
so the surface energy term is the dominant one for $L/b \gg 1$ unless the ratio $\kappa/\sigma b^2$ is very large.

\begin{figure}[htbp]
\begin{center}
   \includegraphics[width=3.0in,angle=270]{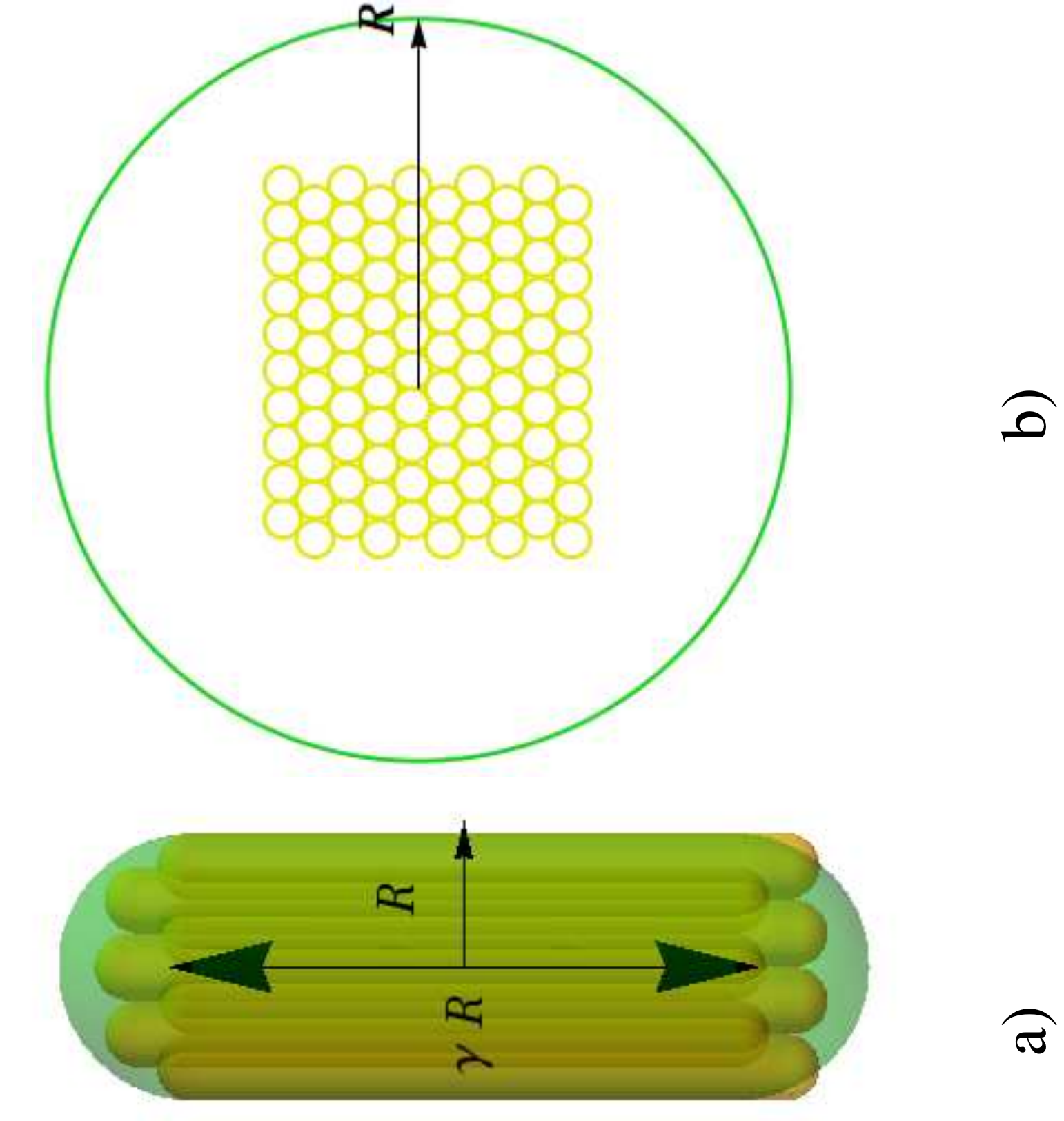} 
\end{center}  
\caption{(Left) Sketch of the DNA condensate as a rod-like structure. (Right)
The hexagonal packing within the cross section of the rod}
\label{fig:fig2}
\end{figure}

Consider now the rod-like structure sketched in Fig. \ref{fig:fig2} as another
possible low energy conformation competing with the toroid. We assume a
spherocylindrical shape for the DNA condensate with circular cross-section of
radius $R$ and length $(\gamma + 2)R$ with $\gamma \ge 0$. The limiting case
$\gamma=0$ again corresponds to a ``globular'' conformation, albeit different
from the previous one. Within this compact structure the polymer chain folds in
such a way that parallel segments are hexagonally packed in the main body of
the structure and there are loops only at the two spherical caps (see
Fig.\ref{fig:fig2}).

As in the previous case, a relation between the length $L$ of the DNA polymer
and the characteristic geometry scale $R$ of the rod-like condensate can be
found hinging upon simple geometrical considerations. The total length $L$ of
the polymer is the sum of two parts: the straight part in the cylinder body
$L_{straight}$, and the loop part in the two spherical caps $L_{loops}$ that
can be estimated as
\begin{eqnarray}
\label{eq:eq19}
L_{straight} = \frac{4\eta \gamma R^3}{b^2}  &\qquad& 
L_{loops}=\frac{16\eta R^3}{3b^2} \ .
\end{eqnarray}
In the above equations, the length is estimated as volume times the packing
fraction of the condensate divided by cross section area of the polymer.
Thus, 
\begin{eqnarray}
\label{eq:eq21}
L &=& L_{straight} + L_{loops} = \frac{4\eta(\gamma+4/3)R^3}{b^2} \ .
\end{eqnarray}
Within the same rationale followed for the toroidal conformation, we assume
each loop to have a constant radius of curvature $R/2$ so that the bending and the
surface energy for the rod-like conformation have the following form

\begin{eqnarray}
\label{eq:eq22}
U_{rod} = \frac{32}{3} \kappa \eta\frac{R}{b} &\qquad
& \sigma S_{rod}= 2\pi \sigma R^2(\gamma+2) . 
\end{eqnarray}

Use of Eq. (\ref{eq:eq21}) then leads to the following forms of the
bending energy and surface energy of the polymer in the rod-like condensate:
\begin{equation}
U_{rod} = \frac{32\kappa}{3} 
\frac{\eta}
{\left[4\eta(\gamma+\frac{4}{3})\right]^{1/3}}
\left(\frac{L}{b}\right)^{1/3} ,
\label{eq:urod}
\end{equation}
\begin{equation}
\sigma S_{rod} = 2\pi (\sigma b^2) \frac{(\gamma+2)}
{\left[4\eta(\gamma+\frac{4}{3})\right]^{2/3}}
\left(\frac{L}{b}\right)^{2/3} .
\label{eq:srod}
\end{equation}
Again, like for the case of toroid condensate, we see that the surface energy
of the rod-like condensate is dominant over the bending energy in the large
$L$ limit.



\subsection{Interaction energy: polyvalent salts}
\label{subsec:interaction}

Up to this point we have not yet considered the actual interactions between neighboring polymer segments but our approach can be easily generalized to include them. 
These interactions can be included on a general poor-solvent level \cite{Depletion}, on the detailed level of explicit electrostatic interactions \cite{Muthukumar2005} or on a phenomenological level based on experimentally determined effective potentials \cite{Todd}. We opt for the latter as the poor-solvent level seems to be too generic while the details of the exact DNA-DNA electrostatic interactions are still incompletely understood \cite{Cherstvy11}. An important reason for sticking to the phenomenological level is that the measured interactions of course contain all the interaction free energy contributions, including the water mediated hydration interaction \cite{Rau84} that do not feature explicitly in model expressions of the poor-solvent or indeed at the electrostatic level.

Let us consider toroidal geometry first and call the contribution of the
interactions between the molecules to the total free energy
\begin{equation}
\label{sec2:eq6b}
U_{int} = U_{int}(\alpha,\eta).
\end{equation}

The form of $U_{int}$ depends on the mode of condensation. In the case of
PSI-condensation it should include the depletion interaction contribution to
osmotic pressure and in the case of the polyvalent counterion condensation it
should include the "strong-coupling" attractive contribution to osmotic
pressure. In general we should have the pairwise interaction potential 
between DNA segments to be {\it ``van der Waals-like''}, but without any
temperature dependence, since temperature is an irrelevant parameter in
condensation, being severely restricted to the interval between the melting of
DNA and freezing of the solvent \cite{footnote}.

\begin{figure}[htpb]
\begin{center}
\includegraphics[width=3.4in]{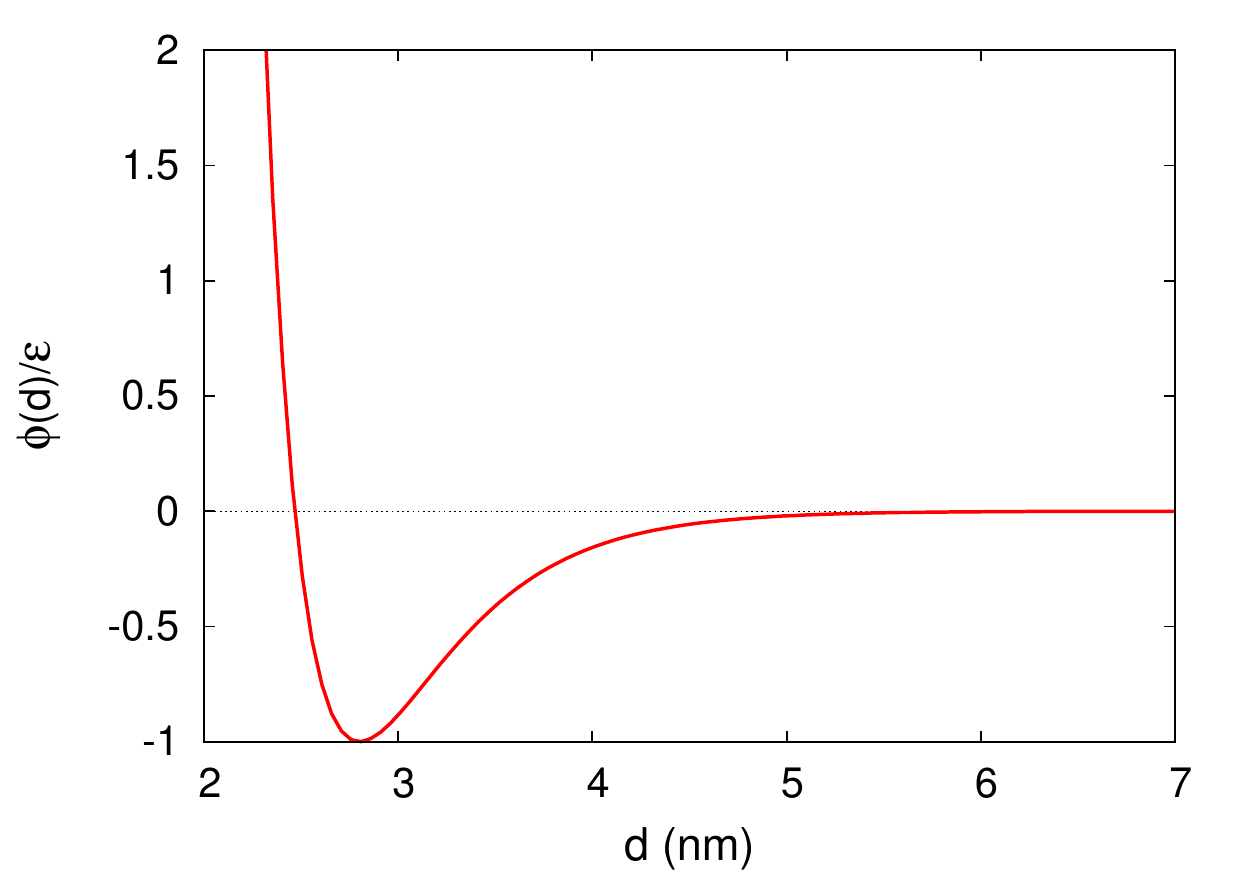}
\end{center}
\caption{Morse potential (Eq. \ref{morsebjko}) plotted with
parameters $d_0=2.8$ nm, $\lambda=0.48$ nm.
}
\label{fig:fig3}
\end{figure}

Let us first discuss DNA condensation in polyvalent salts. In this case the
interaction potential between two parallel neighboring segments of DNA at a
interaxial separation $d$ has been inferred from experiments \cite{Todd} and
has a form accurately described by a Morse potential
\begin{equation}
\widehat{\phi}(d) = \varepsilon \, e^{-2(d-d_0)/\lambda} -
2 \varepsilon \, e^{-(d-d_0)/\lambda} ,
\label{morsebjko}
\end{equation}
per unit length of the interacting straight segments. Here $\varepsilon$
defines the depth of the potential, $d_0$ is the equilibrium interaxial
distance between the molecules at which the potential has a minimum, and 
$\lambda$ characterizes a characteristic length of the potential (see
Fig. \ref{fig:fig3}). Within this model $\varepsilon, d_0$ and $\lambda$
completely parametrize the interactions. For DNA condensation in
[Co(NH$_3$)$_6$]$^{3+}$, a good choice of parameters as inferred from
experiments \cite{Todd} is $d_0 = 2.8$ nm and $\lambda=0.48$ nm, and
$\varepsilon=0.21 k_BT$ per base pair.  Note that a base pair has a
length of $0.34$ nm so 1 $k_BT$/bp corresponds to about 2.94 $k_BT$/(nm).
Upon introducing $\phi(d)=\widehat{\phi}(d) b$, the total interaction energy
between all segments of the chain is then in general given by:
\begin{equation}
U_{int} = \sum_{i<j} \phi(d_{ij}) .
\end{equation}
As this is difficult to evaluate explicitly we introduce an approximation at
this point by considering only the segments that are nearest neighbors and
locally straight. This approximation works fine for short range interactions.
This yields
\begin{equation}
U_{int} \approx N_c \phi(d) ,
\label{eq:uint}
\end{equation}
where $N_c$ is the number of nearest neighbor pairs having distance $d$
between the  polymer segments.  For the assumed hexagonal local packing
symmetry the number of nearest neighbors per segment is 6 inside the condensate
and 4 on its surface.  As the interactions are partitioned between two
neighbors, each segment contributes only half of its interaction energy to the
total energy.  The total number of residues is $L/b$ whereas the number of
residues on the surface of the condensate equals to $S/(bd)$, where $S$
is the condensate's surface area. Thus the total $N_c$ can be estimated as:
\begin{equation}
N_c \approx  3\frac{L}{b} - \frac{S}{bd} .
\label{eq:Nc}
\end{equation}
Note that due to the surface term, $N_c$ depends on both 
$d$ and geometrical parameter $\alpha$ or $\gamma$ depending on the
type of the condensate.

\subsection{Interaction energy: depletion forces}

For the PSI-condensation the effective interactions between polymer segments
include the depletion interaction contribution to osmotic pressure, which is
attractive, stemming from the flexible polymers in solution \cite{Depletion},
and acts on top of a short range repulsive interaction of either electrostatic
or hydration origin \cite{Curop}. The components of the interaction 
energy should thus be a repulsive part and an attractive part that should look
like $-\Pi V_0$, where $\Pi$ is the osmotic pressure of the external polymer
solution and $V_0$ is the overlap of the excluded volume of DNA in the
condensate \cite{Depletion}:
\begin{equation}
U_{int} = U_{repulsion} - \Pi V_0 .
\end{equation}
Let's denote $\delta$ the size of the condensing agent molecule
(in our case the PEG). The overlap volume can be estimated as
\begin{equation}
V_0 = L\pi\left(\frac{b+\delta}{2}\right)^2 -
L\pi\left(\frac{b}{2}\right)^2 \frac{1}{\eta} - \frac{(b+\delta-d)S}{2} ,
\label{eq:V0}
\end{equation}
where the first term corresponds to the excluded volume of the polymer 
in an open conformation, the second term corresponds to the 
volume of the condensate (presumably to be fully excluded from the
osmoticants), the third term corresponds to the excluded
volume of the condensate's surface, and $S$ is the surface area of
the condensate (either toroidal or rod-like). Using Eq. (\ref{sec1:eq3}) one
can rewrite $V_0$ as follows
\begin{equation}
V_0 = L\pi\left(\frac{b+\delta}{2}\right)^2 -
L\pi\left(\frac{d}{2}\right)^2 \frac{1}{\eta_\text{hex}} -
\frac{(b+\delta-d)S}{2} .
\end{equation}
Note that $V_0$ should not be negative which implies that $V_0 \ge 0$
for $d \leq d_c$ with $d_c \approx b+\delta$, and $V_0=0$ otherwise. For a
given $L$ and $\alpha$, the maximum of $V_0$ is obtained at $d=b$.

\begin{figure}[htpb]
\begin{center}
\includegraphics[width=3.4in]{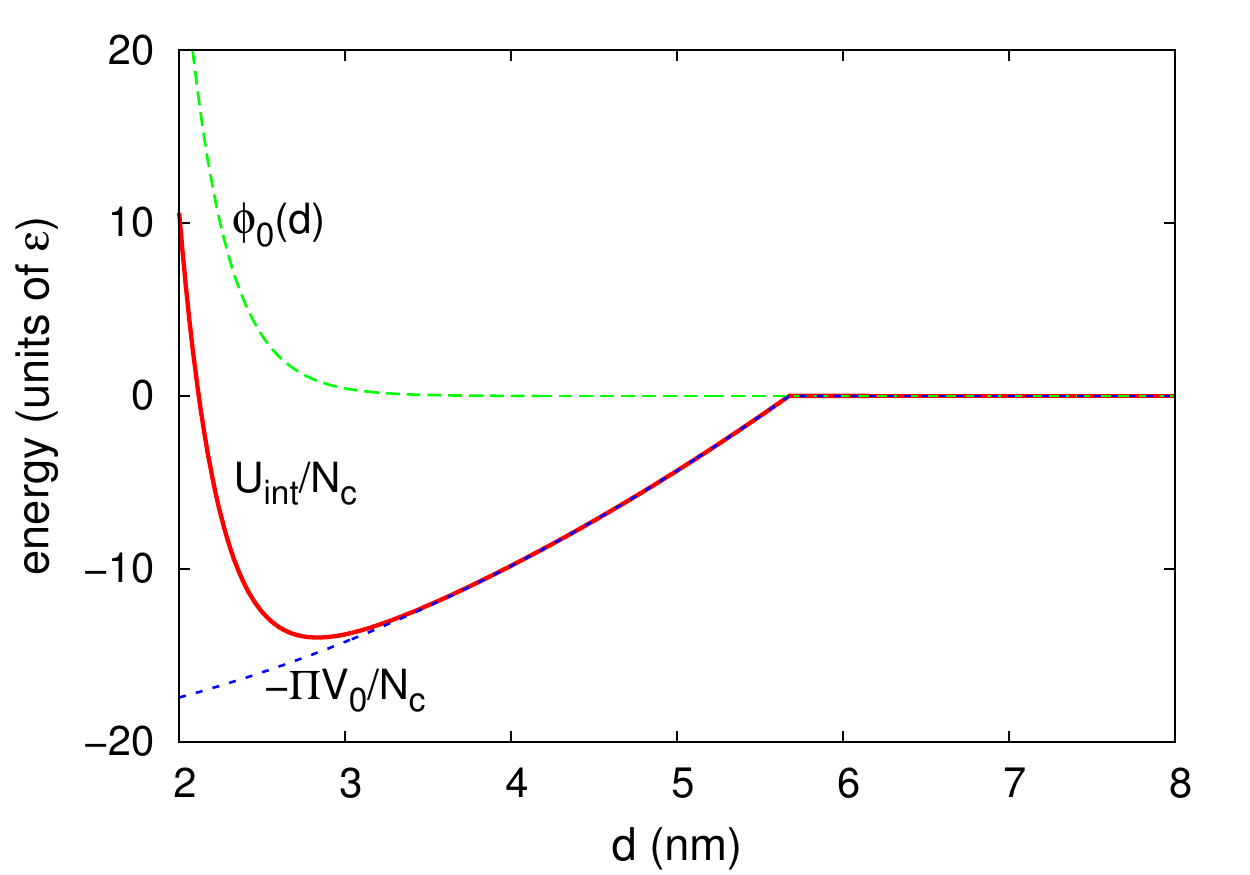}
\end{center}
\caption{Depletion potential (Eq. \ref{depldw}) calculated for toroidal
condensate with parameters for the repulsive potential $d_0=2.8$ nm,
$\lambda=0.48$ nm; the size of osmoticant $\delta=4$ nm; and the osmotic
pressure $\Pi=1\varepsilon/$(nm)$^3$. The dependence on distance $d$ between
nearest neighbor
segments is shown for $U_{int}/N_c$ (solid line), $\phi_0(d)$ (dashed line) and
$-\Pi V_0/N_c$ (dotted line) as indicated.  Note that the position and the
depth of the minimum depends on the value of $\Pi$.}
\label{fig:fig4}
\end{figure}

The repulsive interaction part again contains the energy between all segments
of the chain. Just as before, we again consider only the segments that are
nearest neighbors and locally straight, an approximation that consistently
works for short range interactions. This again yields
\begin{equation}
U_{int} \approx N_c \phi_0(d) - \Pi V_0,
\label{depldw}
\end{equation}
with $\phi_0(d)$ the repulsive part of the Morse potential given
in Eq. (\ref{morsebjko}), i.e.
\begin{eqnarray}
\widehat \phi_0(d) & = & \varepsilon \, e^{-2(d-d_0)/\lambda} , \\
\phi_0(d) &=& b \, \widehat \phi_0(d) .
\label{eq:phi0}
\end{eqnarray}
The osmotic pressure $\Pi$ is given in units of $k_B T/$(unit
length)$^3$. This form of the interaction potential is
routinely seen when compressing DNA with osmoticants such as PEG \cite{RP1}
and is shown on Fig. \ref{fig:fig4}.
The parameters in the repulsive part pertain either to the electrostatic or
hydration interaction, while the osmotic pressure of PEG is known from its
equation of state \cite{Joel}. 

\section{Results}

We now present the results of minimization of the different energy {\sl
Ans\"atze} for a polymer of length $L$. For the toroid-like aggregate with mean
radius $R$ and the thickness $\Delta=\alpha R$ the minimization of
$E_{toroid}(\alpha)$ should be with respect to $\alpha$. For the rod-like
spherocylindrical condensate with circular cross-section of radius $R$ and
length $\gamma R$ the minimization of $E_{rod}(\gamma)$ should be with respect
to $\gamma$. In both cases we first consider the constrained system
with a fixed volume fraction.
We later relax the constraint and consider a system with a soft
interaction potential 
(either the Morse potential or the depletion potential)
that depends on the density of the system. In this case an additional
minimization with respect to the nearest neighbor separation $d$, or
equivalently the density of the system, is in order in both
$E_{toroid}(\alpha,d)$ and $E_{rod}(\gamma,d)$. 

\subsection{Condensates with surface tension}

We first consider an energy {\sl Ansatz} which is composed of the bending energy term
and the surface term only. We assume that the packing fraction $\eta$ is
constant for both the toroidal and rod-like condensate.

For the toroidal condensate, the total energy reads:
\begin{equation}
E_{toroid} (\alpha) = U_{toroid} (\alpha) + \sigma S_{toroid} (\alpha) ,
\label{eq:etot1}
\end{equation}
where $\sigma$ is the surface tension; $U_{toroid}$ and $S_{toroid}$ are given
in Eq. (\ref{sec2:eq6}) and Eq. (\ref{eq:storoid}), respectively. This energy
is dependent both upon the geometry of the torus through the parameter
$\alpha$, the we can seek the optimal configuration by minimizing this energy 
with respect to $\alpha$ at fixed $L$. This minimization leads to the condition
\begin{equation}
\left .
\frac{\partial E_{toroid}}{\partial \alpha} \right |_{\alpha=\alpha^*} = 0
\end{equation}
which yields
\begin{eqnarray}
\label{eq:alpha_star}
\alpha^{*} &=& \frac{1}{8^{3/5}} 
\frac{\pi^{2/5}}{\eta^{4/5}}
\left(\frac{\sigma b^2}{\kappa}\right)^{3/5}
\left(\frac{L}{b}\right)^{1/5}.
\end{eqnarray}
The minimum energy $E_{toroid}^{*}\equiv E_{toroid}(\alpha^{*})$
corresponds to the case where surface and bending energy become comparable and
can be obtained from Eq. (\ref{eq:etot1}) as
\begin{eqnarray}
\label{sec2:eq9}
E_{toroid}^{*}&=& \left(\sigma b^2\right) \frac{5}{2^{7/5}}\frac{\pi^{6/5}}
{\eta^{2/5}} \left(\frac{\kappa}{\sigma b^2}\right)^{1/5}
\left(\frac{L}{b}\right)^{3/5} .
\end{eqnarray} 
Thus the energy of the toroid is minimum when $\alpha=\alpha^*$ if $\alpha^* <
1$ and $\alpha=1$ otherwise. As discussed before, the latter corresponds to a
``globular state''. 
From Eq. (\ref{sec2:eq9}) it is easily seen that both $\alpha^*$ and
$E_{toroid}^*$ are minimum when $\eta$ is maximum. Thus, if $\eta$ is
allowed to vary a further minimization with respect to $\eta$
yields a minimum energy that corresponds to $\eta=\eta_\text{hex}$.

Note that, for any fixed $\kappa/(\sigma b^2) \ge 0$, the ``globule'' will be
the lower energy state as the bending energy is always sub-leading with respect
to the surface energy in the limit  $L/b \gg 1$.  Conversely, for a finite
length $L/b>0$ there will be a critical $\kappa$ beyond which the curvature
term will be dominant. From Eq. (\ref{eq:alpha_star}) this is clearly the case
when $\kappa/(\sigma b^2)$ is large enough so that  $\alpha^* < 1$. This
provides the condition
\begin{equation}
\frac{\kappa}{\sigma b^2} > \frac{1}{8} \frac{\pi^{2/3}}{\eta^{4/3}} 
\left(\frac{L}{b}\right)^{1/3} .
\label{eq:glob}
\end{equation}
This implies a $L^{1/3}$ law for the phase separation between the toroid and
the globule. However, as we will show later in this Subsection, for $\kappa>0$
the globule is always unfavorable against the rod-like condensate, so that in
reality the transition line from a toroid to a globule does not exist in the
ground state phase diagram, if one also takes into account the rod-like
configuration of the condensate.

\begin{figure}[htbp]
\begin{center}
 \includegraphics[width=3.4in]{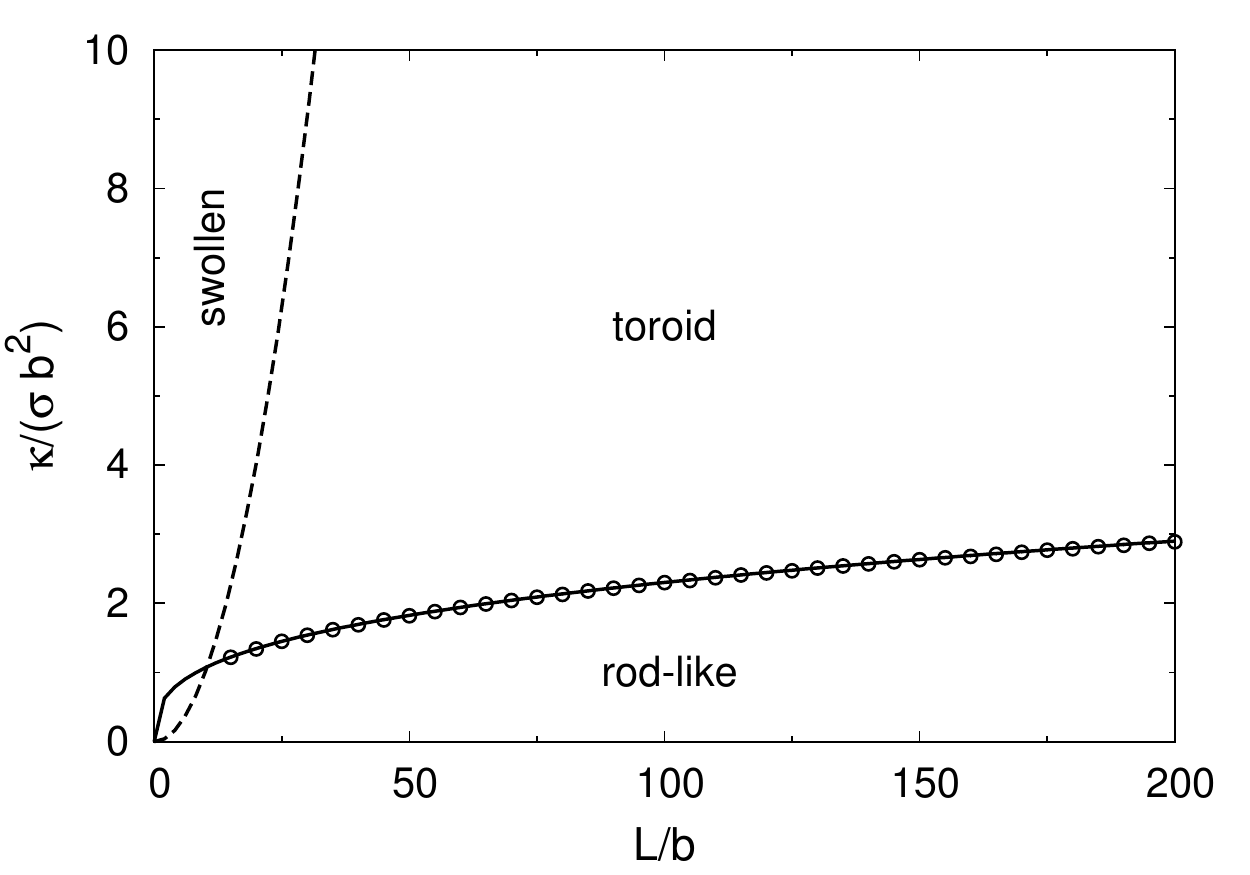} 
\end{center}  
\caption{Expected phase diagram on the basis of the theoretical analysis for
the toroidal, rod-like and swollen phase. The phase diagram was computed 
numerically for the maximum packing fraction $\eta=\eta_\text{hex}$ of the
toroidal and rod-like condensates. The toroidal phase is favorable with
respect to the rod-like counterpart above a certain stiffness, and on the right
of the swollen phase.  The transition line from toroid to swollen (dashed line)
has a $L^2$ dependence as given by Eq. (\ref{eq:quadratic_line}).  The
transition points from toroid to rod-like have been determined numerically from
the condition $E_{toroid}^{*}=E_{rod}^{*}$ (open circles)  and are very well
fitted by a $L^{1/3}$ dependence (solid line).
}
\label{fig:fig5}
\end{figure}

On the other hand, one also requires that the energy of the toroid be smaller
than the energy of the swollen phase which can be approximated with a straight
line conformation:
\begin{equation}
E_{toroid}^{*} < E_{swollen} = \sigma b^2  \pi \frac{L}{b}.
\end{equation}
In the above equation the surface area of the swollen conformation is assumed
to scale as that of a cylinder of length $L$ and diameter $b$. Note that this
conformation has no bending penalty. From Eq. (\ref{sec2:eq9}) 
one obtains
\begin{equation}
\label{eq:quadratic_line}
\frac{\kappa}{\sigma b^2} < \frac{2^7}{5^5} \frac{\eta^{2}}{\pi} 
\left(\frac{L}{b}\right)^{2} .
\end{equation}
The right-hand term of Eq. (\ref{eq:quadratic_line}), yields the dashed line
shown in Fig. \ref{fig:fig5}, above which the bending energy is so large that
the swollen phase is clearly the only optimal conformation.


The energy of the rod-like condensate is given by
\begin{equation}
E_{rod} (\gamma) = U_{rod} (\gamma) + \sigma S_{rod} (\gamma) ,
\label{eq:erod1}
\end{equation}
where $U_{rod}$ and $S_{rod}$ are given in Eq. (\ref{eq:urod}) and Eq.
(\ref{eq:srod}), respectively. $E_{rod}$ depends on the geometry of the
spherocylinder  through the parameter $\gamma$, thus we can obtain the optimal
configuration by minimizing this energy with respect to $\gamma$ at fixed $L$.
This yields
\begin{eqnarray}
\label{eq:eq26}
\frac{(\gamma^* + \frac{4}{3})^{1/3}}{\gamma^*} &=& \frac{3\pi}{4^{7/3}
\eta^{4/3}}
\left(\frac{\sigma b^2}{\kappa} \right) 
\left(\frac{L}{b}\right)^{1/3} .
\end{eqnarray}
The $\gamma^*=0$ solution (the globule) exists only in the limit of $\kappa=0$,
i.e.  non-stiff chain. Thus for semiflexible chains ($\kappa>0$), the rod-like
structure is always energetically favored against the globule and no phase
transition exists  between the rod-like and the globule phases.

The ``ground-state'' energy for the rod-like structure can then be obtained as
$E_{rod}^{*}=E_{rod}(\gamma^{*})$, and compared with the toroid counterpart
$E_{toroid}^{*}$. This provides the full phase diagram in the stiffness-length
plane depicted in Fig. \ref{fig:fig5}, where a transition from a rod-like to a
toroidal conformation is obtained upon increasing $\kappa/(\sigma b^2)$ at a
fixed length.  Remarkably, the transition line from the toroid to a rod-like
condensate follows a $L^{1/3}$ law as revealed by numerical data (Fig.
\ref{fig:fig5}).  The exact location of the intermediate swollen phase,
appearing in Fig. \ref{fig:fig5} for very short lengths, is outside of the
range of applicability of our analysis and it would require further more
specific analysis.

\subsection{Monte Carlo simulations}

We can now compare the scenario obtained by analytical approximations with that
obtained from detailed simulations. To this aim, we have implemented a
standard NVT Monte Carlo simulations on a bead-stick model, where the
DNA chain is modeled as a collection of $N$ consecutive monomers represented
by impenetrable hard spheres of radii $R_{HS}=b/2$ that are tangent to each
other, while the non-consecutive monomers interact via a square-well potential
of range $R_{int}=1.3 b$. We have studied systems of lengths $L$ up to 96
monomers
and stiffness $\kappa$ up to 50$\epsilon$, where $\epsilon$ is the depth
of the square-well potential. An extensive search for the ground states has
been performed in order to construct a phase diagram in the $\kappa$-$L$ space.

\begin{figure}[htpb]
\begin{center}
\includegraphics[width=3.4in]{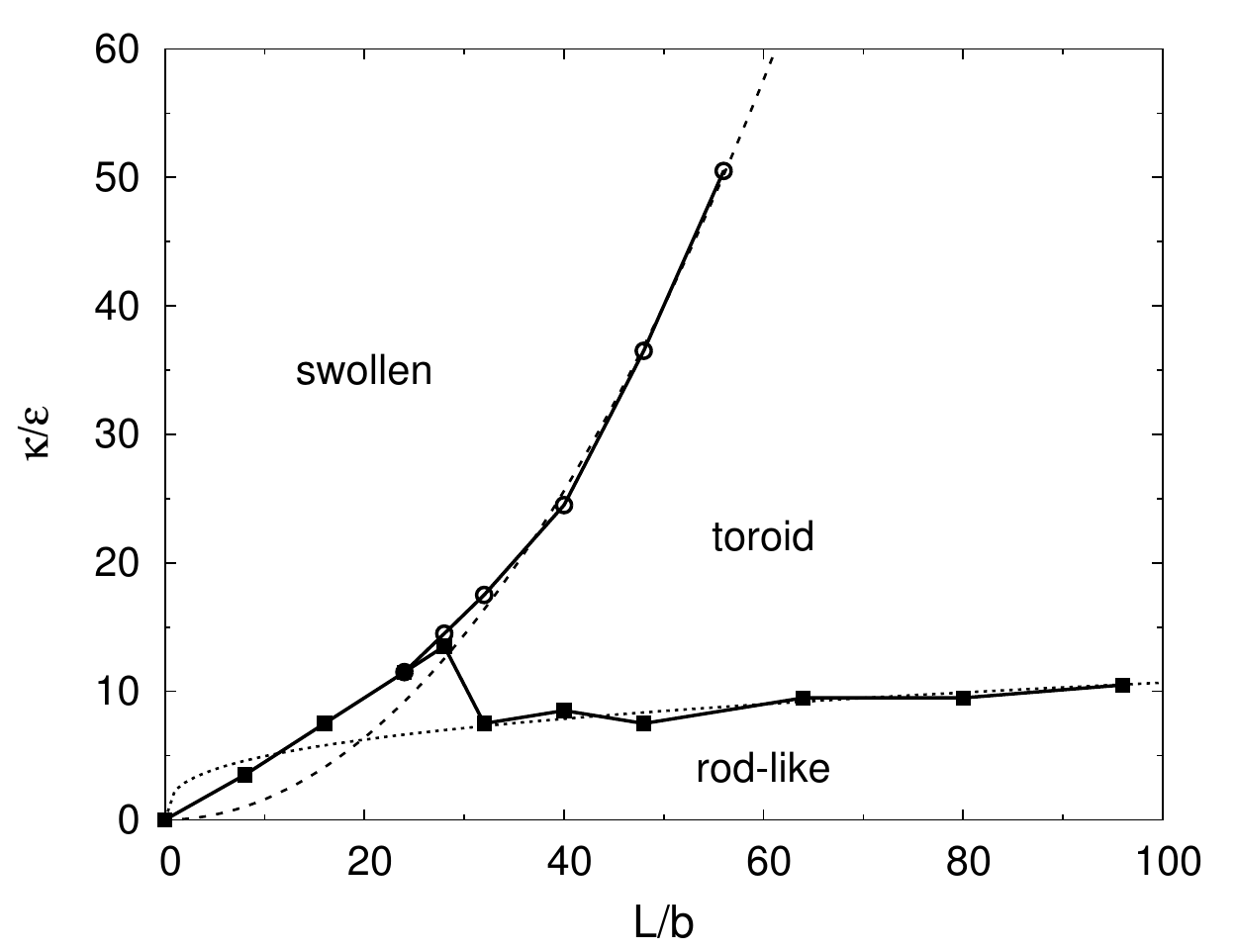}
\end{center}
\caption{
Phase diagram obtained from Monte Carlo simulations displaying toroid and
rod-like ground state energies.  The range of the square-well interaction
was selected to be $R_{int}=1.3 b$, and $R_{HS}=b/2$ is the radius of each
monomer. Underlying broken lines represent a fit of the $(L/b)^{1/3}$
(dotted line) and $(L/b)^2$ (dashed line) dependences found in Fig.
\ref{fig:fig5}.
}
\label{fig:fig6}
\end{figure}

\begin{figure}[htpb]
\begin{center}
\includegraphics[width=3.4in]{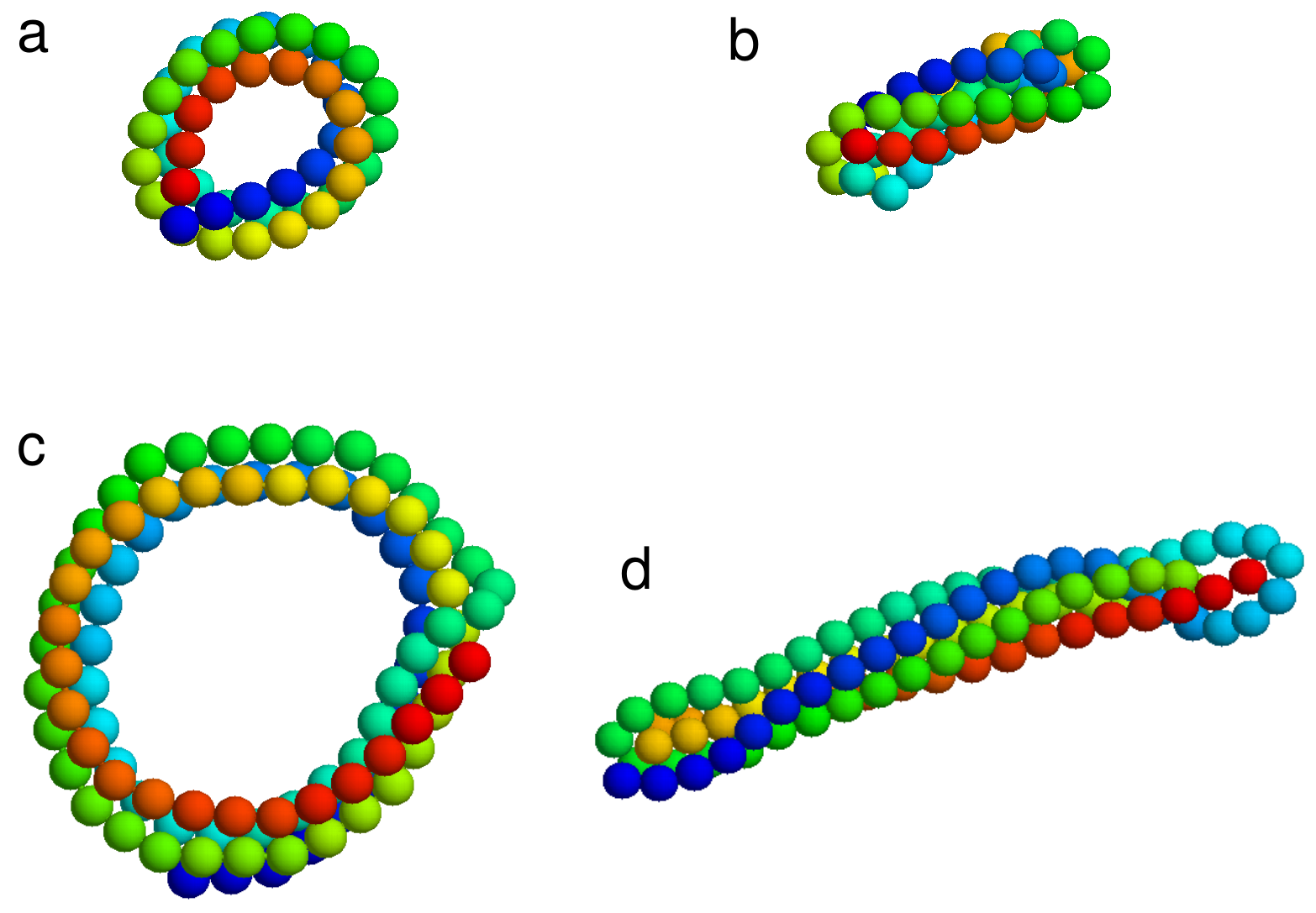}
\end{center}
\caption{Snapshots of the toroidal and rod-like configurations obtained in simulations
of the bead-stick model.  The conformations shown are the lowest energy
conformations for
$N=48$ and $\kappa=10\epsilon$ (a), 
$N=48$ and $\kappa=6\epsilon$ (b), 
$N=96$ and $\kappa=11\epsilon$ (c), 
$N=96$ and $\kappa=9\epsilon$ (d). 
}
\label{fig:fig7}
\end{figure}

The simulations are carried out with standard pivot and crankshaft move sets
and the Metropolis algorithm for move acceptance. A parallel tempering scheme
with 16 replicas is implemented to efficiently equilibrate the system and to
obtain the ground state and low energy conformations at low temperatures.
Consistent results are typically obtained after $1 \div 10\times 10^9$ MC
steps per replica depending on the chain length.

The resulting phase diagram shown in Fig. \ref{fig:fig6} is in remarkable
agreement with Fig. \ref{fig:fig5}, thus confirming the soundness of our
analytical theory. We also present snapshots illustration of toroidal and rod-like
conformations  in Fig. \ref{fig:fig7}.

Very recently, a molecular dynamic study by Lappala and Terentjev appeared
\cite{Lappala13} where they also observed a transition from rod-like to
toroidal condensate above a well defined  persistence length, in full agreement
with our results.

\subsection{Condensates with soft interactions: polyvalent salts}

We now proceed to include the soft interactions between the polymer segments
in the condensate instead of the surface tension, as anticipated earlier. As we
shall see, there will be some general consequences that are independent of the
specific functional form of the interaction potential pointing towards a
universality of the condensation phenomenon in semiflexible polymers.

For the toroidal case the energy {\sl {\sl Ansatz}} is 
\begin{eqnarray}
E_{toroid}\left(\alpha, \eta \right) &=&  U_{toroid} (\alpha,\eta)   
+ U_{int}(\alpha, \eta) \nonumber \\
&=&  U_{toroid} (\alpha,\eta) - \frac{\phi(d)}{bd} S_{toroid} (\alpha,\eta) + 
\nonumber \\
& & + 3\phi(d)\frac{L}{b} \ ,
\end{eqnarray}
where where $U_{int}(\alpha,\eta)$ given in Eq. (\ref{eq:uint}) is the
interaction part depending on the density of the molecules, and
$\phi(d)$ is the Morse potential given in Eq. (\ref{morsebjko}). 
The explicit forms of $U_{toroid} (\alpha,\eta)$ and $S_{toroid} (\alpha,\eta)$
are given in Eq. (\ref{sec2:eq6}) and Eq. (\ref{eq:storoid}), respectively.
Note that $\eta$ depends on $d$ by Eq. (\ref{sec1:eq3}) so that
these two parameters are equivalent.
Analogously for the
rod-like condensate case the energy {\sl {\sl Ansatz}} is 
\begin{eqnarray}
E_{rod}\left(\gamma, \eta \right) &=&  U_{rod} (\gamma,\eta)   
+ U_{int}(\gamma, \eta) \nonumber \\
&=&  U_{rod} (\gamma,\eta) - \frac{\phi(d)}{bd} S_{rod} (\gamma,\eta) + 
\nonumber \\
& & + 3\phi(d)\frac{L}{b} \ ,
\end{eqnarray}

For a fixed value of density, or equivalently $d$, one can find a minimum of
$E_{toroid}\left(\alpha, \eta \right)$ with respect to $\alpha$. This gives 
\begin{eqnarray}
\label{eq:alpha}
\alpha^{*} &=& \frac{1}{8^{3/5}} 
\frac{\pi^{2/5}}{\eta^{4/5}}
\left(\frac{\sigma' b^2}{\kappa}\right)^{3/5}
\left(\frac{L}{b}\right)^{1/5} ,
\end{eqnarray}
where $\sigma' = - \frac{\phi(d)}{bd}$ and $\eta=\eta_{\text{hex}}
\left(\frac{b}{d}\right)^2$. Note that we look only for solution of $\alpha$ in
the range of [0,1]. So $\alpha^* < 0$ corresponds to $\alpha^*=0$ (swollen
conformation) and $\alpha^*>1$ corresponds to $\alpha^*=1$ (globule
conformation).

In the case of the rod-like condensate we get analogously the equation
for $\gamma^*$:
\begin{equation}
\label{eq:gamma}
\frac{(\gamma^* + \frac{4}{3})^{1/3}}{\gamma^*} = \frac{3\pi}{4^{7/3}\eta^{4/3}}
\left(\frac{\sigma' \, b^2}{\kappa} \right) 
\left(\frac{L}{b}\right)^{1/3} .
\end{equation}
If $\sigma' < 0$, there is no solution of $\gamma^* > 0$, the minimum of
$E_{rod}$ is obtained at $\gamma^*=\infty$ (swollen conformation). If $\sigma'
> 0$, a solution $\gamma^* > 0$ always exists if the right hand side is finite.

In order to calculate the phase diagram, we need to compare the minimum energy
of the toroidal condensate with that of the rod-like condensate. The analysis
is pretty similar to that in the previous section with $\sigma'$ playing the
role of surface tension, except that we have now the interaction energy
$\phi(d)$ between polymer segments (and so is $\sigma'$) depending on the
interchain distance $d$ between nearest neighbors. Thus, the minimization of
the energy should be done also with respect to $d$, or equivalently to $\eta$.
It is convenient to do the minimization numerically for both the toroidal and
the rod-like condensates. 

\begin{figure}[htpb]
\begin{center}
\includegraphics[width=3.4in]{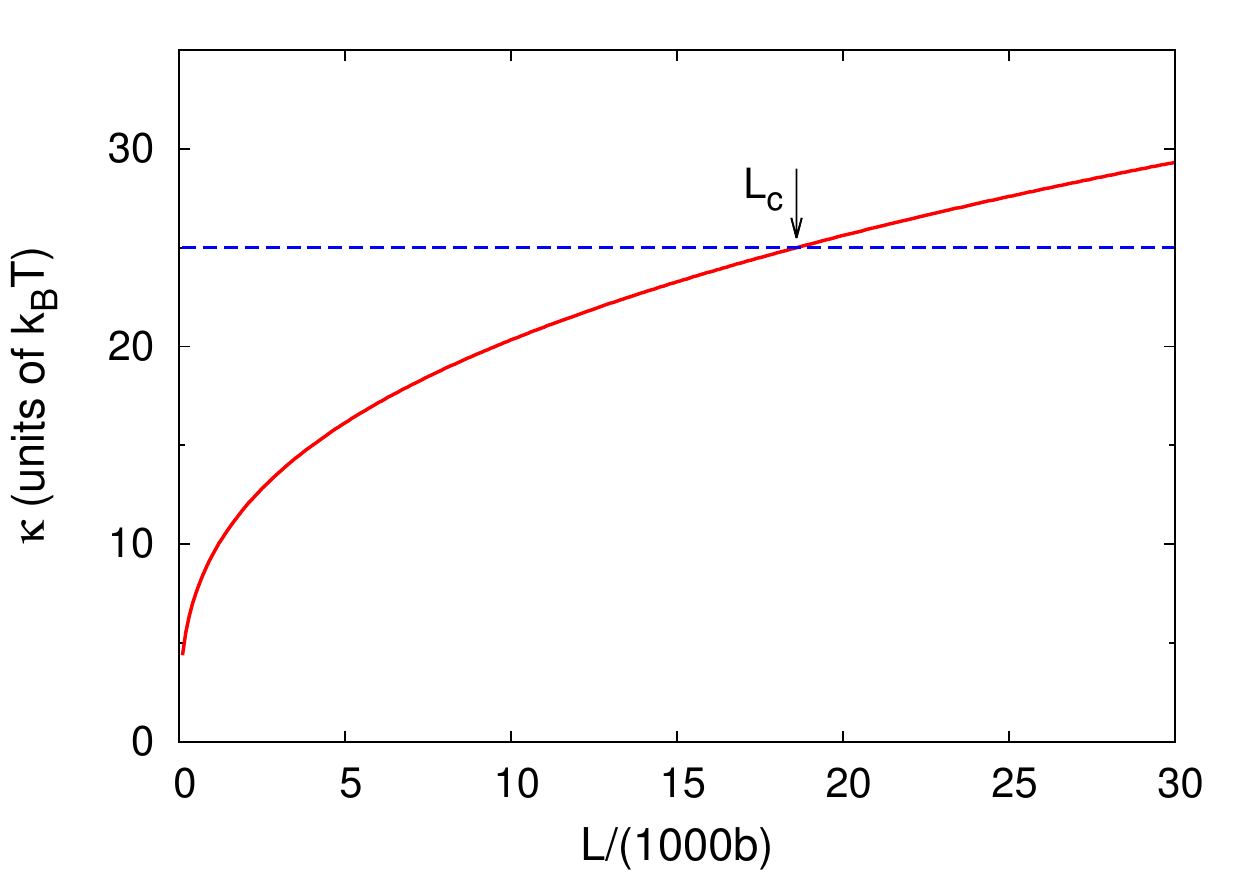}
\end{center}
\caption{Separation between the toroid phase and the rod-like phase
as predicted by our theoretical model
with the Morse interaction potential. The transition line was
calculated numerically and is fitted very well by a $L^{1/3}$ law.
The depth of the interaction potential
is chosen to be $\varepsilon=0.21 k_BT$/bp.  The horizontal line
corresponds to DNA stiffness, $\beta \kappa = l_p/b = 25$,
the DNA diameter $b=2$ nm and persistence length $l_p=50$ nm.
The cut-off length for the toroid is
$L_c=18.6\times10^3b$ or equivalent to 109.4 kbp.
}
\label{fig:fig8}
\end{figure}

Fig. \ref{fig:fig8} shows the phase diagram in the stiffness-length plane
obtained for realistic parameters of the Morse potential. The latter (shown in
Fig. \ref{fig:fig3}) is calculated with $b=2$ nm (the diameter of DNA), $d_0 =
2.8$ nm (the equilibrium interaxial distance between neighboring base pairs),
$\lambda=0.48$ nm (the width of the potential), $\varepsilon = 0.21 k_BT$/bp
(the depth of the potential). The phase separation line behaves like $L^{1/3}$,
as in the model with only the bending energy term and the surface term. Using a
realistic parameter for DNA stiffness, $\kappa=25 k_BT$, it is found that the
critical length $L_c$ for the toroid is equivalent to 109.4 kbp. For
$L>L_c$ the ground state becomes the rod-like structure.

\begin{figure}[htpb]
\begin{center}
\includegraphics[width=3.4in]{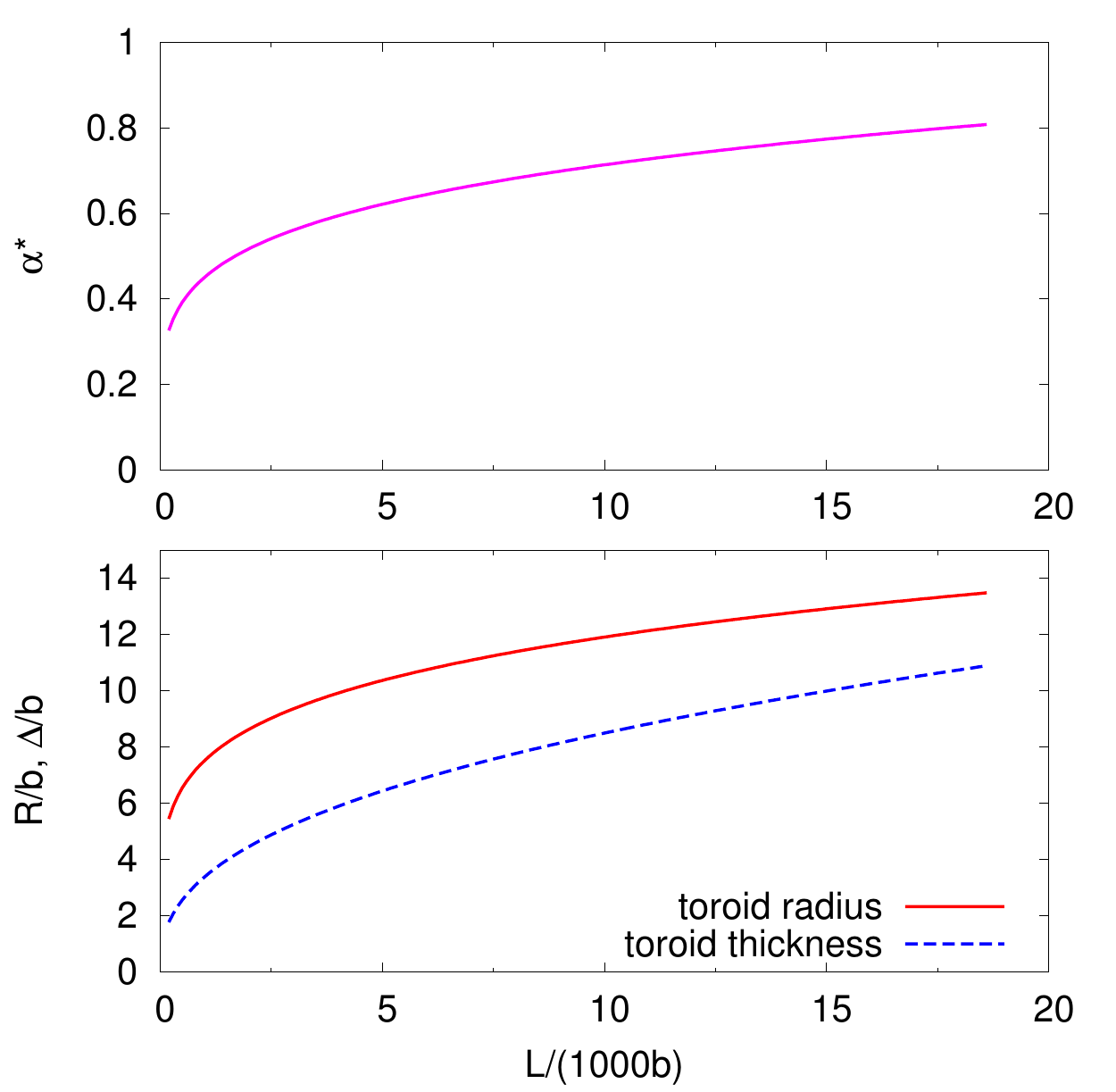} 
\end{center}
\caption{(Top) Numerical solution of Eqs. (\ref{eq:alpha}) as a function of the
chain length. (Bottom) The corresponding growth of toroid radius, $R$, and
toroid thickness, $\Delta$ with chain length, $L$, that can be fitted by a
$L^{1/5}$ law for the radius and a $L^{2/5}$ law for the thickness,
respectively. The dependences are shown for $L<L_c$.
The data are obtained for the model with Morse interaction
potential with realistic parameters for DNA with $\epsilon=0.21 k_B T$/bp.
}
\label{fig:fig9}
\end{figure}

Fig. \ref{fig:fig9} (Top) shows the numerical solutions for the toroid
geometrical parameter $\alpha^{*}$ as function of the chain length $L$ for
realistic parameters of DNA. We find that $\alpha^*$ grows with the chain
length like $L^{1/5}$ up to a maximum value $\alpha^*_c \approx 0.807$ at the
phase separation ($L=L_c$). We have also found that the value of $\alpha^*_c$
does not depend on the chain stiffness and interaction energy parameter.
Further in Fig. \ref{fig:fig9} (Bottom), the dependence of the
toroid radius $R$ and thickness $\Delta=\alpha R$ on the total contour length
$L$ of the DNA is also depicted. The toroid radius is found to grow with the
chain length like $L^{1/5}$ whereas its thickness grows like $L^{2/5}$.

\begin{figure}[htpb]
\begin{center}
\includegraphics[width=3.4in]{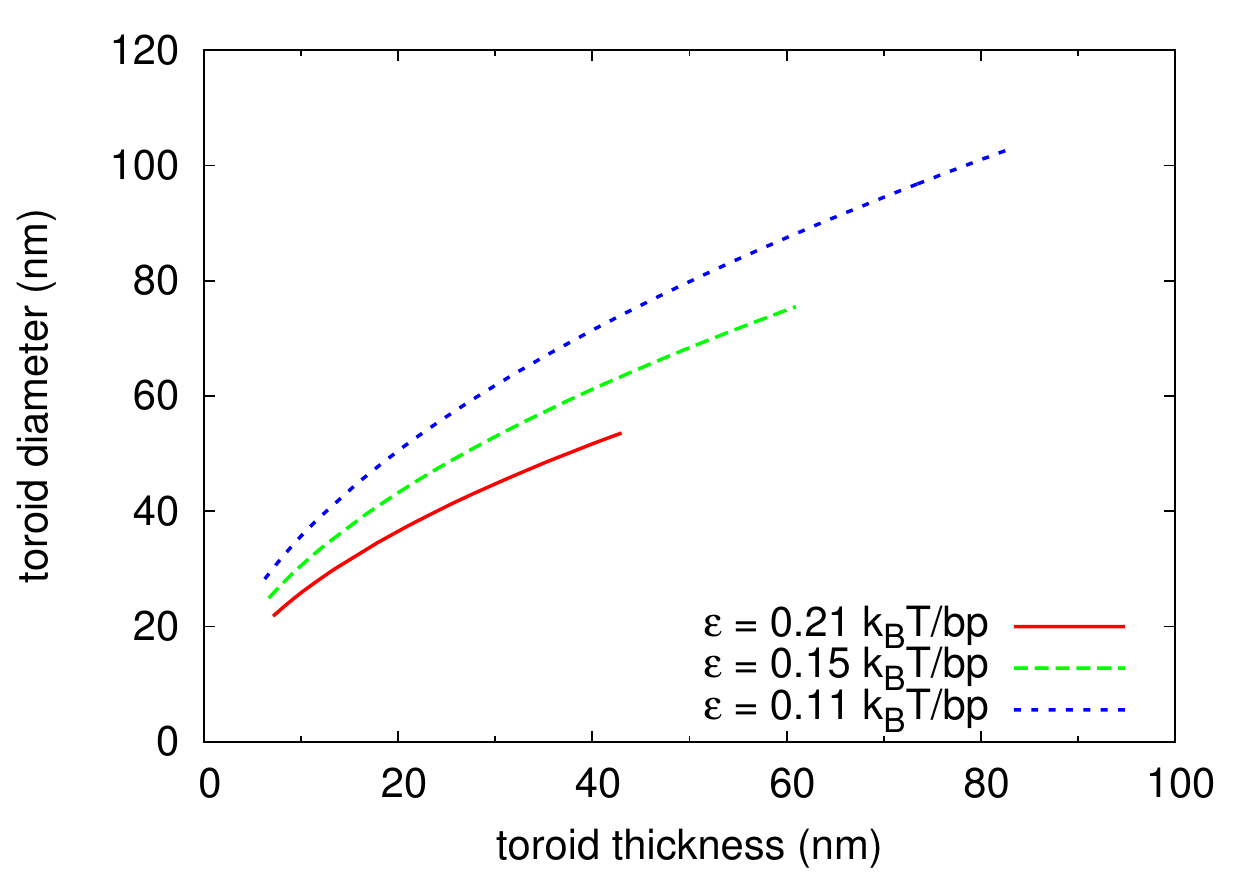}
\end{center}
\caption{Toroid thickness ($=2 \alpha R$) vs. toroid diameter ($=2 R$) as predicted by our theoretical
model for several values of interaction energy parameter $\varepsilon$ as
indicated. The toroid thickness and diameter are calculated for length $L<L_c$
at which the toroid is the ground state. The DNA parameters used here were
$b=2$ nm, $d_0=2.8$ nm, $\lambda=0.48$ nm, and $\kappa=25 k_B T$.
}
\label{fig:fig10}
\end{figure}

An interesting finding of our theoretical analysis concerns a prediction of the
toroid thickness's dependence on the toroid radius, as illustrated in Fig.
\ref{fig:fig10}, and is in very good agreement with the experimental findings of
Conwell et al. \cite{Conwell03}. Our results clearly shows that the ratio of
the toroid diameter to toroid thickness strongly depends on the solvent
condition as given by the energy parameter $\varepsilon$. The poorer the
solvent, the smaller is the toroid diameter.  Our model could be used to
infer the interaction energy between base pairs from experimental data of toroid
sizes for various solvent conditions.
Furthermore, for a given value of $\epsilon$, our model indicates that (toroid diameter) $ \propto $ (toroid thickness)$^{1/2}$. This
relation could be verified experimentally by direct measurements of toroid dimensions.

\subsection{Condensates with soft interactions: depletion potential}

We consider now the third energy {\sl Ansatz} in which the energy of the condensate
is composed of the bending energy and the interaction energy due to osmotic
depletion. For the toroidal condensate, the energy takes the following form:
\begin{equation}
E_{toroid} (\alpha,d) = U_{toroid} (\alpha,d) + N_c \phi_0(d)
- \Pi V_0 (\alpha,d) ,
\end{equation}
where $N_c$ is given in Eq. (\ref{eq:Nc}) and $V_0$ is given in Eq.
(\ref{eq:V0}). It is straightforward to show that for a given $d$, the energy
is minimized at the value of $\alpha^*$ given by
\begin{eqnarray}
\label{eq:alphadep}
\alpha^{*} &=& \frac{1}{8^{3/5}} 
\frac{\pi^{2/5}}{\eta^{4/5}}
\left(\frac{\sigma' b^2}{\kappa}\right)^{3/5}
\left(\frac{L}{b}\right)^{1/5} ,
\end{eqnarray}
with $\sigma' = \Pi\frac{b+\delta-d}{2}- \frac{\phi_0(d)}{bd}$. 
Similarly for the rod-like condensate, the equation for $\gamma^*$ is the same
as Eq. (\ref{eq:gamma}) with the new $\sigma'$ as in the above equation.
Note that in order for the toroid to be stable, $\sigma'$ should be positive,
this leads to the following condition for the osmotic pressure:
\begin{equation}
\Pi \geq \min_{d<b+\delta} \ \frac{2\phi_0(d)}{bd(b+\delta-d)} = \Pi_{min},
\end{equation}
where the minimum is taken over all $d$ satisfying $d<d+\delta$. This minimum
value of the osmotic pressure strongly depends on the diameter of the
osmoticant. For example, for the repulsive interaction given in Eq. 
(\ref{eq:phi0}) with realistic parameters for DNA, $\Pi_{min} \approx
2.2$ $k_BT$/(nm)$^3$ and $0.025$ $k_BT$/(nm)$^3$ for $\delta=1$ nm 
and 2 nm, respectively. So with a small osmoticant it is much harder to
condense DNA than with larger ones. Note that an effective diameter of $2$ nm
would correspond to PEG-6000 as an osmoticant.

\begin{figure}[htpb]
\begin{center}
\includegraphics[width=3.4in]{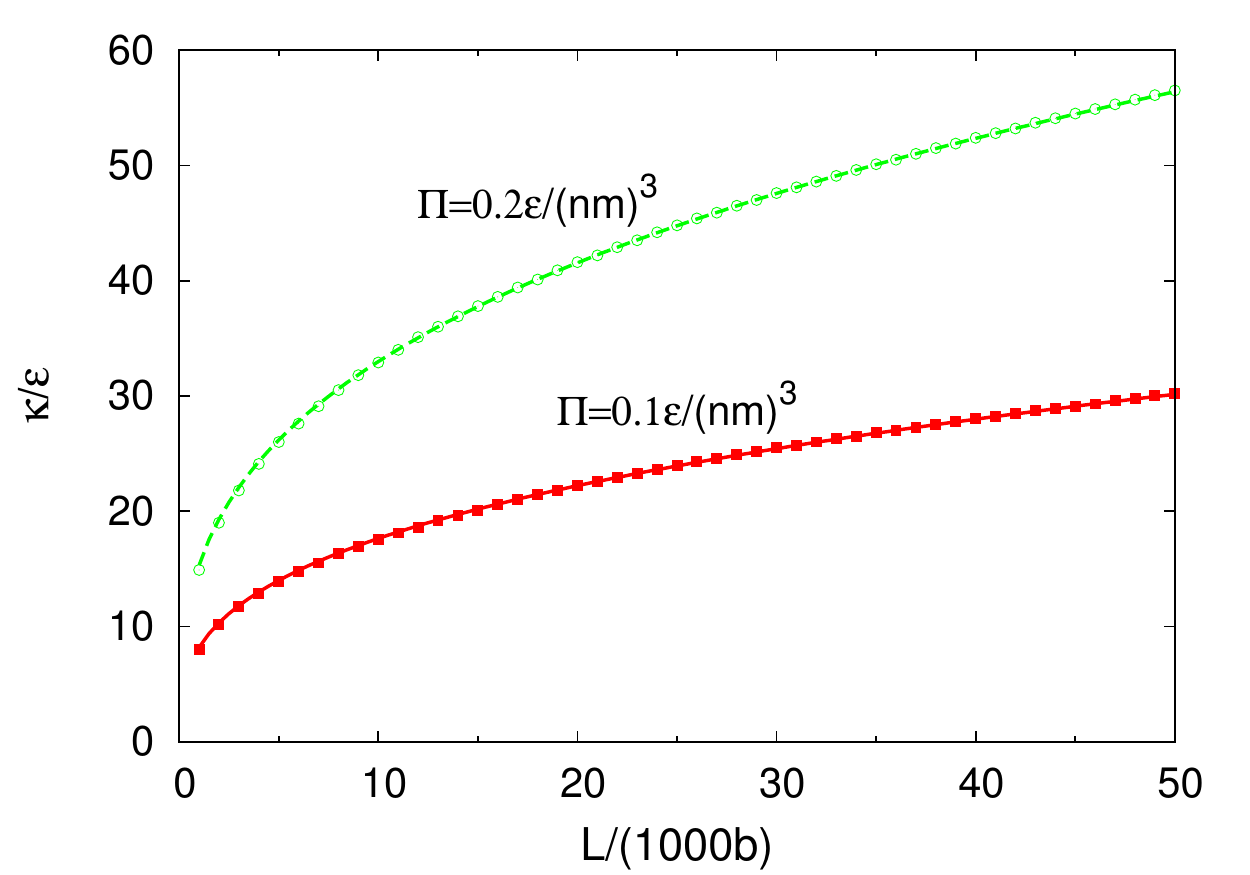}
\end{center}
\caption{Phase separation between the toroid and the rod-like structure in the
$\kappa$-$L$ diagram for two values of osmotic pressure $\Pi$ as indicated.
The numerical data (squares and circles) can be fitted very well with
$L^{1/3}$ (solid and dashed lines) for both cases.
}
\label{fig:fig11}
\end{figure}

\begin{figure}[htpb]
\begin{center}
\includegraphics[width=3.4in]{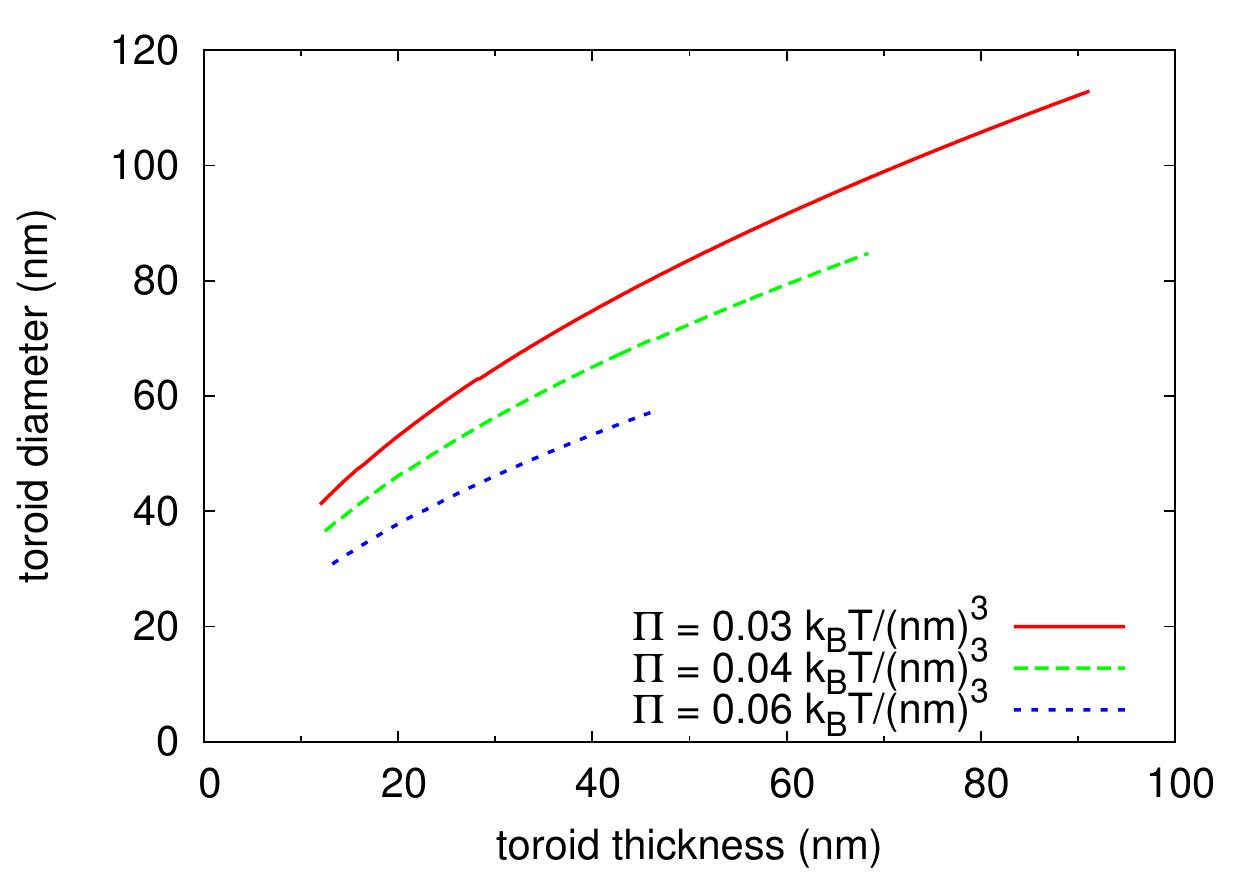}
\end{center}
\caption{
Dependence of toroid diameter on toroid thickness in the model with
depletion potential. The data are shown for three values of osmotic
pressure $\Pi$ as indicated. 
}
\label{fig:fig12}
\end{figure}

Figs. \ref{fig:fig11} and \ref{fig:fig12} show the phase diagram and the
dependence of toroid diameter on toroid thickness for the case with depletion
potential. Note that phase separation line well follows the $L^{1/3}$ law,
and the stronger osmotic pressure, the smaller sizes of toroids are formed.
The results are qualitatively similar to those obtained with the
Morse potential and the hard-core interaction case.  Therefore the detailed
form of the interaction potential, while being important for the equation of
state of DNA \cite{Strey99}, has only a small effect on the location of the
transition lines and on the length dependence of the various phases themselves.

\subsection{Nonlinear elasticity: stiff chain with an elastic threshold}

Up to now we have only considered the variations in the interaction potential
on the phase diagram of the condensate. Inspired by the tube model of a polymer
\cite{Maritan00,Banavar03,Poletto08}, we finally also consider a variation in
the form of the bending energy of the polymer. Specifically we intend to
introduce modifications in the $\theta$ dependence of the curvature energy.

Assume first that the curvature energy vanishes for $\theta \leq \theta_0$  and
becomes $\infty$ otherwise. The ensuing radius of curvature is in this case also limited
by a corresponding lower bound $R_0$
\begin{equation}
\label{sec4:eq1}
1 - \cos \theta = \frac{b^2}{2 R^2} < 1 - \cos \theta_0 \ ,
\end{equation}
or
\begin{equation}
\label{sec4:eq2}
R > \frac{b}{\sqrt{2\left(1-\cos\theta_0\right)}} \equiv R_0 \ .
\end{equation}

Under  these assumptions, the polymer will try to make very large turns (i.e.
large radius of curvature) in order to have $U=0$. Hence a rod-like structure,
where short turns are unavoidable, will then be consequently highly
unfavorable, as further elaborated below. 

When the condition (\ref{sec4:eq2}) is met, the total energy coincides with
surface energy only, so that
\begin{eqnarray}
\label{sec4:eq3}
E_{toroid} (\alpha) & = & \sigma S_{toroid} =  \sigma b^2 4 \pi^2 \alpha 
\left(\frac{R}{b}\right)^2 \nonumber \\
 &=& \left(\sigma b^2\right) \frac{\pi^{4/3}}{\alpha^{1/3} \eta^{2/3}} \left(\frac{L}{b}\right)^{2/3} ,
\end{eqnarray}
in the same form as given in Eq. (\ref{eq:storoid}). From the above
equation one can see that for a given length $L$ the minimum energy is obtained
when both $\eta$ and $\alpha$ are at their maximum values, i.e.
$\eta=\eta_\text{hex}$ and $\alpha=\alpha^*$. On the other hand,
from Eq. (\ref{eq:Rb}) it is clear that for a given $L$, $\alpha$ is
maximum when $R$ is minimum, thus one obtains the following equation
for $\alpha^*$:
\begin{equation}
\frac{L}{b} = 8\pi\eta_\text{hex} (\alpha^*)^2 \left(\frac{R_0}{b}\right)^2 .
\end{equation}
The above equation has a solution $\alpha^* < 1$ only if
\begin{equation}
L < L_0 \equiv 8\pi\eta_\text{hex} b \left(\frac{R_0}{b}\right)^2 .
\label{eq:LL0}
\end{equation}
In other words, the toroid can be a minimum energy configuration only when
$L<L_0$. For $L\geq L_0$ the minimum energy would be a globule ($\alpha^*=1$).
Because there is no bending energy, the globule would be always favored
compared to the rod-like conformation, having the lowest surface energy. It
then follows that within this model, only two phases are possible: the toroid
phase for length $L < L_0$ and the globule phase for $L > L_0$ (see Fig.
\ref{fig:fig11}). Here $L_0$ is the smallest length necessary to form a globule
and can be determined from Eqs. (\ref{eq:LL0}) and (\ref{sec4:eq2}) as
\begin{eqnarray}
\label{sec4:eq5}
\frac{L_0}{b} &=& \frac{2^{3/2}\pi \eta_{\text{hex}}}{\left(1-\cos\theta_0\right)^{3/2}} .
\end{eqnarray}

\begin{figure}[htpb]
\begin{center}
\includegraphics[width=3.4in]{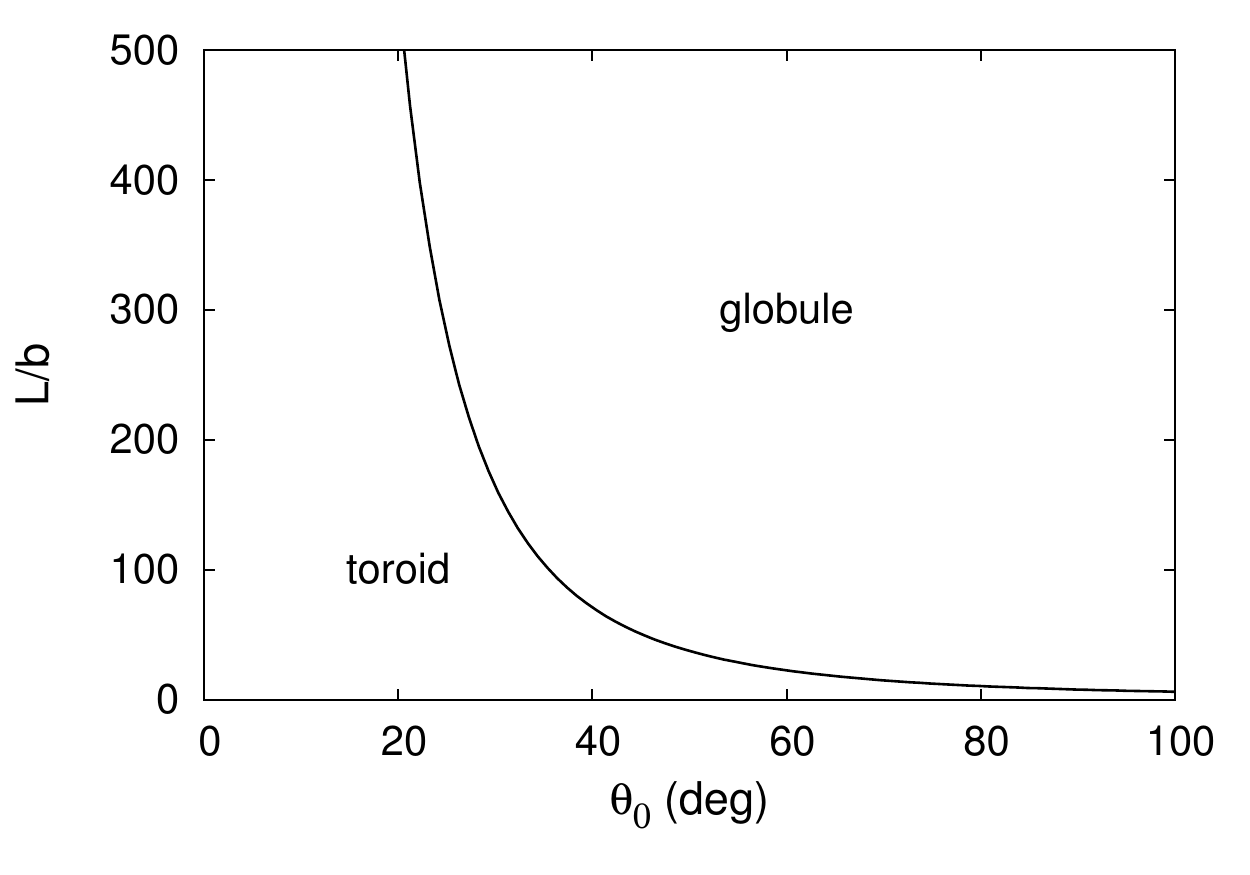}
\end{center}
\caption{
Ground state phase diagram for a stiff chain with a threshold on
angle $\theta$. Solid line is where $L=L_0$.}
\label{fig:fig13}
\end{figure}

For $L<L_0$ it is easy to show that
\begin{equation}
\label{sec4:alpha}
\alpha^* = \left(\frac{L}{L_0}\right)^2 ,
\end{equation}
and
\begin{equation}
\label{sec4:etormin}
E_{toroid}^{*} = \left(\sigma b^2\right) 
\frac{\pi^{4/3}}{\eta_{\text{hex}}^{2/3}}
\left(\frac{L_0}{b}\right)^{1/6} 
\left(\frac{L}{b}\right)^{1/2} .
\end{equation}
We have also checked by extensive simulations that the ground state of a
self-attracting polymer with a threshold on angle $\theta$ is  a toroid (data
not shown) for the chain length $L<L_0$ in accordance with the phase diagram
shown in Fig. \ref{fig:fig13}.

\section{Conclusions}

We have studied the phase diagram of semi-flexible polymers with
self-attraction as a function of the stiffness and the contour length of the
chain. This is a prototypical model for DNA condensation, a problem with a long
history that is still not completely understood.

We have combined analytical arguments with extensive Monte Carlo simulations, to study the competition between toroidal and rod-like configurations as candidates for the ground state of the condensate at increasing stiffness of the polymer molecules. As the stiffness increases, we find an increasing tendency for the polymers to achieve a nematic alignment in order to minimize the bending energy. This typically favours first a transition from a globular to a rod-like conformation, and then a further transition to a toroidal conformation that is then the most stable one at sufficiently large stiffness values. This scenario appears to be rather robust irrespective of the details of the interaction potential between the segments of the molecules, provided that the polymer chain is sufficiently long.

In the stiffness-length plane, we furthermore found a robust $L^{1/3}$ dependence of the transition line between the rod-like and the toroidal phases, that  can be simply explained within our analytical approach. An additional $L^{1/5}$ dependence was also observed for the toroid radius. While the existence of some of these scaling laws has been known for some time, see the work of Grosberg and Zhestkov Ref. \onlinecite{history}, a comprehensive analysis combining an analytical approach with numerical simulations has been, to the best of  our knowledge, still missing. Our work fills this gap.

Finally, our conclusions are in good agreement also with very recent molecular dynamics simulations \cite{Lappala13} that studied the dynamics of a polymer chain collapse in poor solvents as a function of the chain flexibility. We assessed the importance of the exact form of the interaction potential, as extracted from experiments, while constructing models for polyvalent salt condensed DNA as well as polymer and salt induced DNA condensation. We also investigate the effects of non-linear elasticity within the specific model of a tube model of a polymer that has important consequences on the phase diagram of the semi-flexible chain.

\section{acknowledgments}
AG acknowledges funding from  PRIN-COFIN2010-2011 (contract 2010LKE4CC).
TXH and NTTN thank for support from National Foundation for Science and Technology Development (NAFOSTED Grant 103.01-2010.11). RP acknowledges support from the Agency for Research and Development of Slovenia (ARRS grant P1- 0055(C)).



\begin{thebibliography}{99}
\bibitem{footnote} This is not true in the case of $\rm Mn^{2+}$ condensed DNA that does show a large temperature effect. See Ref.
\cite{Rau92}.
\end{thebibliography}
\end{document}